\documentclass[preprint,showpacs,preprintnumbers,amsmath,amssymb]{revtex4}
\usepackage{graphicx}% Include figure files
\usepackage{dcolumn}% Align table columns on decimal point
\usepackage{bm}% bold math
\usepackage{amssymb}
\usepackage{amsmath}
\usepackage{latexsym}

\begin{document}

%\draft
\title{
Derivation of general Maxwell type equations for open quantum systems
}
\author{
A.V. Khugaev$^{1,2}$
G.G. Adamian$^{1}$,  N.V. Antonenko$^{1}$
}
\affiliation{
$^{1}$Joint Institute for
Nuclear Research, 141980 Dubna, Russia\\
$^{2}$Institute of Nuclear Physics,
100214 Tashkent, Uzbekistan
}

\date{\today}
%\maketitle

\begin{abstract}
Using   equations of motion with the anisotropic dissipative term for
quantum particle and quantum-mechanical commutation rules,
 the general Maxwell-type differential equations are
derived. The direct modifications of the well-known Maxwell   equations  due to the
medium effects (openness of the system) are  discussed.

\end{abstract}


\maketitle
\section{Introduction}
A Maxwell equations have been derived as a
 mathematical reflection of the already established an empirical
 facts. In the macroscopic quantum electrodynamics, they have been expressed as a
 relationships between electric ${\bf E^0}$ and magnetic ${\bf H^0}$ field vectors, describing an
 electromagnetic field.
  The properties of the medium (matter) are taken into account only by means of material equations.
  Since the latter is not always possible, such a description is not universal \cite{M4}.

Using the equations of motion
\begin{equation}
m\ddot{x}_j=F^0_j({\bf x},\dot{{\bf x}},t)
\label{e1}
\end{equation}
of the non-relativistic  point-particle of mass
$m$ with the Cartesian   position operator
$x_j$, $j=1,2,3$,
and
the quantum mechanical commutation relations
\begin{equation}
[x_{j},x_{k}]=0,
\label{e2}
\end{equation}
\begin{equation}
m[x_j,\dot{x}_k]=i\hbar\delta_{jk},
\label{e3}
\end{equation}
%[x_{j},x_{k}]=0\nonumber\\
%where $j,k=1,2,3$ are a spatial indexes.
%a first couple of the Maxwell equations
the Lorentz force law ($\epsilon_{jkl}$ is the Levi-Civita antisymmetric tensor)
\begin{equation}
F^0_j=E^0_j+\epsilon_{jkl}\dot{x}_kH^0_l
\label{e4}
\end{equation}
and two
Maxwell equations
\begin{equation}
\nabla_j H^0_j=0,
\label{e5}
\end{equation}
\begin{equation}
\dot{H^0_j}+\epsilon_{jkl}\nabla_{k}E^0_l=0
\label{e6}
\end{equation}
have been derived in Ref. \cite{Dyson}.
%Here, $\epsilon_{jkl}$ is the Levi-Civita antisymmetric tensor.
As explicitly  demonstrated,
${\bf E^0}({\bf x},t)$
and ${\bf H^0}({\bf x},t)$
are the vector functions of the coordinate ${\bf x}$ and time $t$ but not the velocities. As  shown in Ref. \cite{Hughes},
the particle motion in the non-inertial frame with the Coriolis like forces or in the weak
gravitational field also satisfies the constraints (\ref{e1})--(\ref{e3}).  The velocity-dependent forces
are not limited to electromagnetic ones, and the commutation relations (\ref{e2}) and (\ref{e3})
are also determined by the equations of motion (\ref{e1})  \cite{Hughes}.
%
%
% One of the main question  discussed in Ref. \cite{Dyson} has been
% a question about inconsistency of initial assumptions of the paper and
% obtained final result where author started from the pure non-relativistic
% assumptions and comes to the Maxwell equations which are relativistically
% invariant.
The generalization of this  approach \cite{Dyson,Hughes} to the relativistic case
has been done in Ref. \cite{Japan}. The connection of the classical equations of motion and Maxwell
electromagnetic equations in  elegant manner has been explicitly shown in Ref. \cite{Naresh1}.
Note also  the derivation of the Lorentz force  law in Ref. \cite{Berard} based on the commutation rules.

In Refs.  \cite{Dyson,Hughes,Japan,Naresh1,Berard},
it has been not taken into account that   systems by
their   nature are open systems \cite{M1,M2,M3,M4,M5,M6}.
In this case, it is necessary to explicitly take into consideration
the environmental or medium effects in the   equations of motion.
In the present paper, we try to solve this problem.
%that is, to take into account the interaction of the system or particle with the environment.
%
% In our short notes we will give some remarks to the derivation
%of the first couple of the Maxwell equations and after that we will try to
%find in the same manner a possible generalization of these equations with
%taking into the account an influence of the medium in terms of the dissipation
%of initial energy of moving particle. But before it we want discuss
%the derivation itself from the physical point of view and give a brief
%motivation of the paper.
%
%
%In the present paper, the main question under the our consideration is how it is
% possible to change the equation of motion, remaining the other assumptions
% of the paper\cite{Dyson} and using the same technique for the obtaining some more
% general expressions for the Maxwell like equations, than it was obtained in\cite{Dyson}.
In the addition to the Lorenz force $F^0_j$, the motion of the test particle (open system) is under
 the influence of an additional force $f_j$ induced by physical medium.
% and which are taken in the form: $f_j=-\lambda_{jk}({\bf x}) \dot{x}_k$, where $\lambda_{jk}({\bf x})$
% is some general tensor field, depending from coordinates and
This force effectively describes   the interaction of the moving particle with medium
and has the experimental justification.

\section{General consideration}
%
%Let's take a relationships from the paper\cite{Dyson} in the same form and
%notations, but with some small corrections as it was above declared:
Let us consider the general equations of motion
\begin{equation}
m\ddot{x_j}=F_j({\bf x},\dot{{\bf x}},t)-\lambda_{jk}({\bf x})\dot{x}_k,
\label{e1n}
%\label{e7}
\end{equation}
%\begin{equation}
%[x_{j},x_{k}]=0
%\label{e2}
%%\label{e8}
%\end{equation}
%\begin{equation}
%m[x_j,\dot{x}_k]=i\hbar\delta_{jk}
%\label{e3}
%%\label{e9}
%\end{equation}
where the motion of the test particle is under
 the influence of an additional dissipative type force
 $f_j=-\lambda_{jk}({\bf x}) \dot{x}_k$  induced by physical medium.
Here, $\lambda_{jk}({\bf x})=\lambda_{(jk)}({\bf x})+\lambda_{[jk]}({\bf x})$  is the general tensor field,
where  $\lambda_{(jk)}$
and   $\lambda_{[jk]}$ are the  symmetrized and anti-symmetrized  tensors, respectively.
Note that these tensors depend on
the coordinate ${\bf x}$.
In general, Eq. (\ref{e1n}) should also include the fluctuating forces, leading to
a stochastic dynamics. For simplicity, we consider the average dynamics of the test particle.

% which are neglected.
Taking the total time derivative of  Eq. (\ref{e3}), we obtain
\begin{equation}
[\dot{x}_j,\dot{x}_k]+[x_j,\ddot{x}_k]=0.
\label{e2n}
%\label{e8}
\end{equation}
Substituting (\ref{e1n}) in Eq. (\ref{e2n}), we find
\begin{equation}
[x_j,F_k]=\frac{i\hbar}{m}\Biggl(\lambda_{(jk)}+\Biggl\{\frac{im^2}{\hbar}[\dot{x}_j,\dot{x}_k]-\lambda_{[jk]}\Biggr\}\Biggr)=\frac{i\hbar}{m}\Biggl(\lambda_{(jk)}-\epsilon_{jkl}H_l \Biggr).
\label{e3n}
\end{equation}
Because the  term in curly brackets in Eq. (\ref{e3n}) is anti-symmetrized with respect to $j\to k$ and  $k\to j$,
we introduce, without loss of generality, the vector "magnetic" field
\begin{equation}
H_l=-\frac{1}{2}\epsilon_{jkl}\Biggl(\frac{im^2}{\hbar}[\dot{x}_j,\dot{x}_k]-\lambda_{[jk]}\Biggr)=H^0_l+\frac{1}{2}\epsilon_{jkl}\lambda_{[jk]}.
\label{e5n}
%\label{e10}
\end{equation}
%or
%\begin{equation}
%\frac{i\hbar}{m}\epsilon_{jkl}H_l=\frac{i\hbar}{m}\lambda_{[kj]}+m[\dot{x_j},\dot{x_k}].
%\label{e6n}
%\label{e9}
%\end{equation}
%Here, $\epsilon_{jkl}$  is the Levi-Civita symbol.
%Employing the Jacobi identity
%\begin{equation}
%[x_l,[\dot{x_j},\dot{x_k}]]+[\dot{x_j},[\dot{x_k},x_l]]+[\dot{x_k},[x_l,\dot{x_j}]]=0
%\label{e7n}
%%\label{e11}
%\end{equation}
%and Eq. (\ref{e3}),
%we obtain
%\begin{equation}
%[x_l,[\dot{x_j},\dot{x_k}]]=0.
%\label{e8n}
%%\label{e12}
%\end{equation}
%%
Expressing $[\dot{x}_j,\dot{x}_k]$ from Eq. (\ref{e5n}) and
substituting  into the Jacobi identity
\begin{equation}
[x_l,[\dot{x}_j,\dot{x}_k]]=0,
\label{e8n}
%\label{e12}
\end{equation}
we derive
\begin{equation}
\epsilon_{jkp}[x_l,H_p]=[x_l,\lambda_{[jk]}]=0\to [x_l,H_p]=0,
\label{e9n}
\end{equation}
which means that $H_p$ only depends on  ${\bf x}$ and $t$.
The Jacobi identity
\begin{equation}
\epsilon_{jkl}[\dot{x}_l,[\dot{x}_j,\dot{x}_k]]=0
\label{e15}
\end{equation}
together with   Eq. (\ref{e5n}) implies
\begin{equation}
[\dot{x_l}, H_l]=\frac{1}{2}\epsilon_{jkl}[\dot{x}_l,\lambda_{[jk]}]
\label{e16}
\end{equation}
which is equivalent to
\begin{equation}
div{\bf H}=\frac{1}{2}\epsilon_{jkl}\nabla_l\lambda_{[jk]}.
\label{e19}
\end{equation}
One can assign the role of "magnetic" charge density $\rho_m$  to
the term in the right hand side of Eq. (\ref{e19}). So,
 the "magnetic" charge density $\rho_m$   is a possible source of  the static "magnetic" field
by analogy with  the electric charge density $\rho_e$  as a source of  the static electric field.
Here,  the "magnetic" charge  ("Dirac monopole" \cite{Di}) is related to the heterogeneous antisymmetric tensor field $\lambda_{[jk]}$.
When $\lambda_{[jk]}$ is a homogeneous field  and, correspondingly,  $\nabla_l\lambda_{[jk]}=0$,
we again obtain Eq. (\ref{e5}). In the magnets, the  magnetic  charge is related with the
magnetization ${\bf I}$:  $\rho_m=-{\rm div}{\bf I}$ \cite{Vons}.
%As well-known,
The well-known  Maxwell equations   assert
that  $\rho_m=0$ and there are no other sources of   magnetic fields, except electric currents.

Substituting now the vector field
\begin{equation}
F_j=E_j+\epsilon_{jkl}\dot{x_k}H_l
\label{e10n}
%\label{e12}
\end{equation}
into Eq. (\ref{e3n}), we obtain
\begin{equation}
[x_j,E_k]=\frac{i\hbar}{m}\lambda_{(jk)}.
\label{e11n}
%\label{e13}
\end{equation}
%this expression differ from the according
%commutator in\cite{Dyson} at $\lambda_{(jk)}\neq 0$.
The  vector "electric" field
\begin{equation}
E_j=E^0_j+\lambda_{(jk)}\dot{x}_k
\label{e12n}
\end{equation}
follows from   Eqs. (\ref{e2}) and  (\ref{e11n}).
As seen from Eqs. (\ref{e5n}) and (\ref{e12n}),
the  vector "electric" and "magnetic" fields
contain the individual property of the medium,
"resistance" or "conductivity". At ${\bf E^0}=0$,
Eq. (\ref{e12n}) is reduced to the material-type equation
(e.g., the Ohmic law in the case of  electromagnetic forces).
Using Eq. (\ref{e12n}), one can derive the Maxwell type equation:
\begin{equation}
div{\bf E}=\rho_e+\nabla_j\lambda_{(jk)}\dot{x}_k,
\label{e14n}
\end{equation}
where  $\rho_e=div{\bf E^0}$ is an analog of   the "electric" charge density.

Taking  total time derivative of the vector "magnetic" field ${\bf H}$ (\ref{e5n}),
\begin{equation}
\frac{\partial H_l}{\partial t}+\dot{x}_m\frac{\partial H_l}{\partial x_m}
=\frac{1}{2}\epsilon_{jkl}\dot{x}_m\nabla_m\lambda_{[jk]}-\frac{im^2}{\hbar}\epsilon_{jkl}[\ddot{x}_j,\dot{x}_k],
\label{e20}
\end{equation}
and making some tedious but  simple algebra, we
obtain the following  Maxwell-type equation
\begin{eqnarray}
\frac{\partial H_l}{\partial t}=-\epsilon_{lkj}\frac{\partial E_j}{\partial {x}_k}-
\frac{1}{m}\biggl\{\lambda_{jj}H_l-\lambda_{ql}H_q-\lambda_{[lk]}H_k-
\epsilon_{jkl}\lambda_{jp}\lambda_{[pk]}\biggr\}\nonumber \\
-\epsilon_{jkl}\dot{x}_p\nabla_k\lambda_{jp}
+\frac{1}{2}\biggl\{\epsilon_{jkl}\dot{x}_m\nabla_m +\epsilon_{jkp}\dot{x}_l\nabla_p \biggr\}\lambda_{[jk]}.
\label{e26}
\end{eqnarray}
This equation is the generalized law of "electromagnetic" induction
and more complicated than the corresponding Maxwell equation.
In the case of $\lambda_{jk}\equiv 0$, we re-derive all results of Refs. \cite{Dyson,Hughes}.
%However, it is
%a more interesting to analyze the difference between our results and results of Refs. \cite{Dyson,Hughes}.
If the $\lambda_{jk}$ are constants, Eq. (\ref{e26}) is transformed into
\begin{eqnarray}
\frac{\partial H_l}{\partial t}=-\epsilon_{lkj}\frac{\partial E_j}{\partial {x}_k}-
\frac{1}{m}\biggl\{\lambda_{jj}H_l-\lambda_{ql}H_q-\lambda_{[lk]}H_k-
\epsilon_{jkl}\lambda_{jp}\lambda_{[pk]}\biggr\}.
\label{e29}
\end{eqnarray}
In   particular case  of  the symmetric tensor field $\lambda_{jk}=\lambda ({\bf x})\delta_{jk}$,  Eq. (\ref{e26}) is simplified
as
\begin{equation}
\frac{\partial H_l}{\partial t}=-\epsilon_{lkj}\frac{\partial E_j}{\partial {x}_k}-
\frac{2\lambda}{m}H_l-\epsilon_{ljk}\dot{x}_j\nabla_k\lambda
\label{e27}
\end{equation}
or
\begin{equation}
\frac{\partial{\bf H}}{\partial t}=-{\rm rot}{\bf E}-
\frac{2\lambda}{m}{\bf H}-\dot{{\bf x}}\times {\bf\nabla}\lambda
\label{e28}
\end{equation}
which is similar to one for the ferromagnetic materials, in which $\lambda$
is proportional to the magnetic conductivity (inverse to the magnetic viscosity) \cite{Vons}.
At constant $\lambda$, we have
\begin{equation}
\frac{\partial{\bf H}}{\partial t}=-{\rm rot}{\bf E}-
\frac{2\lambda}{m}{\bf H}.
\label{e30}
\end{equation}

Taking  total time derivative of vector "electric" field ${\bf E}$ (\ref{e12n}),
\begin{equation}
\frac{\partial E_l}{\partial t}+\dot{x}_m\frac{\partial E_l}{\partial x_m}
=\frac{\partial E^0_l}{\partial t}+\dot{x}_m\frac{\partial E^0_l}{\partial x_m}+\dot{x}_i\nabla_i\lambda_{(lk)}+\lambda_{(lk)}\ddot{x}_k,
\label{e32}
\end{equation}
we  obtain the following Maxwell-type equation
%(the generalized "Ampere's" law)
\begin{eqnarray}
\frac{\partial E_l}{\partial t}=\epsilon_{lkj}\frac{\partial H_j}{\partial {x}_k}-j_{l}+\dot{x}_i\nabla_i\lambda_{[lk]}\dot{x}_k
+\frac{\lambda_{(lk)}}{m}\biggl\{E_k+\epsilon_{lkj}\dot{x_l}H_j-\lambda_{kl}\dot{x}_l\biggl\}.
\label{e33}
\end{eqnarray}
Here, we employ that
\begin{eqnarray}
\frac{\partial E^0_l}{\partial t}=\epsilon_{lkj}\frac{\partial H^0_j}{\partial {x}_k}-j_{l},
\label{e333}
\end{eqnarray}
where ${\bf j}$ is an analog of  the "electric" current density.
Eq. (\ref{e33}) is the generalized "Ampere's" law.

Thus, in the general case for electromagnetic forces, instead of  the Maxwell   and  material  equations,
  more complicated equations should be used:
  a closed system of coupled equations of motion (\ref{e1n})
  [or more general - the quantum Langevin equations or the corresponding quantum diffusion equation
  by taking also into account the quantum  fluctuations]  for the charge particles
  and field equations (\ref{e19}), (\ref{e14n}), (\ref{e26}), (\ref{e33}).
  Note that our definitions for the $H$ and $E$ are different from the standard definitions.
  For example, such type approach can be used to describe electromagnetic processes
  in a fully ionized plasma.

%It should be  noted that similar equations of the Maxwell  type can be derived
%not only for electromagnetic forces, but also for forces of any nature \cite{Hughes}.
%It is possible to assume that the forces in nature are united only by
%the general form of the equations of motion which contain the "electric" and "magnetic" components.

%It can be assumed that only the general form of the equations of motion
%unites all forces in nature.

\section{Conclusions}
Employing an equations of motion for the test quantum particle and quantum-mechanical commutation rules,
we derived the  Maxwell type  differential equations for forces $F_j$ of any nature.
Because
these equations contain the influence of the medium  (openness of the system),
they are more  complicated than  the usual  Maxwell equations for electromagnetic forces.
%
%The Maxwell equations are a closed system of equations for the electric and magnetic field strengths.
% The properties of the medium (matter) are taken into account only by means of material equations.
%  Since the latter is not always possible, such a description is not universal \cite{M4}.
%  In the general case, instead of  the Maxwell   and  material equations,
%  more complicated equations should be used:
%  a closed system of coupled equations of motion (\ref{e1n})
%  [or more general - the corresponding quantum kinetic equation
%  by taking also into account the quantum  fluctuations]  for the charge particle
%  and field equations (\ref{e19}), (\ref{e14n}), (\ref{e26}), (\ref{e33}).
%  For example, such type approach can be used to describe electromagnetic processes
%  in a fully ionized plasma.
%
 As seen,  for the strongly inhomogeneous anisotropic
  systems the effective "magnetic" charge   appears in Eq. (\ref{e19}).
   So, the obtained equations acquire   more symmetrical form,
   where there are  "magnetic" and "electric" charges.
   The "magnetic" charge   is related to the inhomogeneous antisymmetric tensor field $\lambda_{[jk]}$.
   It should be noted that
   magnetic monopole (magnetic charge) has been sought in some materials possessing  strongly
   anisotropic crystal  structure (e.g., the nematic materials) and,
   accordingly, possessing strongly anisotropic dissipative properties.
%   From our point of view, searches in such environments are not meaningless,
%    although until now they have been unsuccessful.
    We also found
    that   the influence of    homogeneous and
    isotropic    medium leads to the field equation (\ref{e30})
    for the ferromagnetic materials with the magnetic conductivity.

Using the  nonrelativistic classical  equations of motion, corresponding to Eqs. (\ref{e1n}),
and the usual commutator-Poisson bracket correspondence,
one can deduce the same results for  the open  classical systems.
It is also possible to assume that the forces in nature are united only by
the general form of the equations of motion
%and field equations
which contain the "electric" and "magnetic" components.

%It should be noted that similar equations of the Maxwell  type can be derived
%not only for electromagnetic forces, but also for forces of any nature \cite{Hughes}.
%It is possible to assume that the forces in nature are united only by
%the general form of the equations of motion which contain the "electric" and "magnetic" components.
%It can be assumed that only the general form of the equations of motion
%unites all forces in nature.

%One can also take   an additional force
% $f_j=-\int^{t}_{0}K_{jk}(t-t^{'})\dot{x}_k(t^{'})dt^{'}$
% induced by   medium  and  construct a complete system of the Maxwell type
%  equations including the non-Markovian effects.
%  Such kind of
%the description will be important when the relaxation
%time of medium are comparable with the specific time
%of perturbation in terms of the medium and moving particle back-responses.

\section*{Acknowledgments}

A.V.K. thanks Bogoliubov Laboratory of Theoretical Physics  (JINR, Dubna)
for their invitation and warm hospitality.
This work was partially supported by  the Russian Foundation for Basic Research (Moscow)   and DFG (Bonn).
The IN2P3(France)-JINR(Dubna) Cooperation
Programme is gratefully acknowledged.

%\section*{References}


\begin{thebibliography}{0}

\bibitem{M4} Yu. L. Klimontovich, {\it Statistical Theory of Open Systems} (Kluwer, Dordrecht, 1995).
%\bibitem{M10} V. V. Dodonov, O. V. Man'ko, and V. I. Man'ko, J. Russ. Laser Res., {\bf 16}, 1 (1995).
%\bibitem{parabola} H.~Hofmann, Phys. Rep. {\bf 284}, 137 (1997).
%\bibitem{Zub}  D. Zubarev, V. Morozov, and G. R\"opke, {\it Statistical Mechanics
%of Nonequilibrium Processes} (Akademie Verlag, Berlin, 1997).
%%, Vol. II.

\bibitem{Dyson} F.J. Dyson,  Am. J. Phys.  {\bf 58}, 209 (1990).

\bibitem{Hughes} R.J. Hughes,   Am. J. Phys.  {\bf 60}(4), 301 (1991).

%\bibitem{AharonovBohm} Y. Aharonov and D. Bohm, {\it Phys. Rev.,} {\bf 115}, No 3, 485 (1959).

%\bibitem{Land2} L.D. Landau and E.M. Lifshitz, Vol.~2, {\it The classical theory of fields}
%(Butterworth-Heinemann, 1984)

%\bibitem{Land8} L.D. Landau, L.P. Pitaevsky and E.M. Lifshitz, Vol.~8,
%{\it Electrodynamics of Continuous Media}(Butterworth-Heinemann, 1984)

%\bibitem{Dirac1} P.A.M. Dirac, {\it Proc. R.Soc., London,} {\bf A113}, 60 (1931).
%\bibitem{Dirac2} P.A.M. Dirac, {\it Phys. Rev.,} {\bf 74}, 817 (1948).

\bibitem{Japan} S. Tanimura,   Ann. Phys.  {\bf 220}, 229 (1992).

\bibitem{Naresh1} P. Singh and N. Dadhich,   Int. J. Mod. Phys. {\bf A16}, 1237 (2001);
  Mod. Phys. Lett.  {\bf A16}, 83 (2001).

\bibitem{Berard} A. Berard, Y. Grandati, and H. Mohbrach,   Phys. Lett.  {\bf A254}, 133 (1999).

\bibitem{M1}  V. V. Dodonov and
V. I. Man'ko, {\it Density Matrices and Wigner Functions of Quasiclassical Quantum Systems}
(Proc. Lebedev Phys. Inst. of Sciences, Vol. {\bf  167}, A. A. Komar, ed.),
Nova Science, Commack, N. Y. (1987).

\bibitem{M2}  K. Lindenberg and B.J. West,
 {\it The Nonequilibrium Statistical Mechanics of Open and Closed Systems}
 (VCH Publishers, Inc., New York,  1990).



\bibitem{M3} A. Isar, A. Sandulescu, H. Scutaru, E. Stefanescu, and W. Scheid,
Int. J. Mod. Phys. E {\bf 3}, 635 (1994).

\bibitem{M5}  U. Weiss, {\it Quantum Dissipative Systems} (Wold Scientific, Singapore,
1999).

\bibitem{M6} V.V. Sargsyan, Z. Kanokov, G.G. Adamian, and N.V. Antonenko,
Phys. Part. Nuclei {\bf 41}, 175 (2010).

\bibitem{Di} P.A.M. Dirac,   Proc. R. Soc.  London  {\bf A113}, 60 (1931);    Phys. Rev.  {\bf 74}, 817 (1948).

\bibitem{Vons} S.V. Vonsowsky, \textit{Magnetism}  (Nauka Publishers, Moscow,  1971).

\end{thebibliography}
\end{document}